\documentclass{aa}  

\usepackage{txfonts}
\usepackage[export]{adjustbox}
\usepackage{subcaption}
\usepackage[colorlinks,linkcolor=blue,anchorcolor=blue,citecolor=blue]{hyperref}
\usepackage{xcolor}
\usepackage{upgreek}
\usepackage{amsmath}
\usepackage{aas_macros}
\usepackage{amssymb}
\usepackage[thinc]{esdiff}

\usepackage{siunitx}
\DeclareSIUnit\gauss{G}

\newcommand{\uint}{$\upmu\rm Jy\,beam^{-1}$}
\newcommand{\umag}{$\upmu\rm G$}

\begin{document} 

    \title{CHANG-ES XXXIV: a 20\,kpc radio bubble in the halo of the star-forming galaxy NGC\,4217}


   \author{V. Heesen\inst{1}
          \and
          T. Wiegert\inst{2}
          \and
          J. Irwin\inst{3}
          \and
          R. Crocker\inst{4}
          \and 
          A. Kiehn\inst{1}
          \and 
          J.-T. Li \inst{5}
          \and
          Q. D. Wang\inst{6}
          \and
          M. Stein\inst{7}
          \and
          R.-J. Dettmar\inst{7}
          \and
          M.~Soida\inst{8}
          \and 
          R. Henriksen\inst{3}
          \and
          L. Gajovic\inst{1}
          \and 
          Y. Yang\inst{5}
          \and
          M. Br\"uggen\inst{1}
         }

   \institute{Hamburg University, Hamburger Sternwarte, Gojenbergsweg 112, 21029 Hamburg, Germany\\
              \email{volker.heesen@hs.uni-hamburg.de}
        \and
        Instituto de Astrofısica de Andalucıa (IAA-CSIC), Glorieta de la Astronomıa, 18008, Granada, Spain
        \and
        Department of Physics, Engineering Physics, \& Astronomy, Queens University, Kingston, ON K7L 3N6, Canada
        \and
        Research School of Astronomy and Astrophysics, Australian National University, Canberra 2611, A.C.T., Australia
        \and 
        Department of Astronomy, University of Massachusetts, North Pleasant Street, Amherst, MA 01003-9305, USA
        \and
        Purple Mountain Observatory, Chinese Academy of Sciences, 10 Yuanhua Road, Nanjing 210023, China
        \and
        Ruhr University Bochum, Faculty of Physics and Astronomy, Astronomical Institute (AIRUB), 44780 Bochum, Germany
        \and 
        Astronomical Observatory, Jagiellonian University, ul. Orla 171, 30-244 Krak\'ow, Poland
        }

   \date{Received date / Accepted date}
   
 
  \abstract
   {Cosmic rays may  be dynamically very important in driving large-scale galactic winds. Edge-on galaxies give us an outsider's view of the radio halo, which shows the presence of extra-planar cosmic-ray electrons and magnetic fields.}
   {We present a new radio continuum imaging study of the nearby edge-on galaxy NGC\,4217 in order to study the distribution of extra-planar cosmic rays and magnetic fields. We both observe with the {\it Jansky} Very Large Array (JVLA) in the $S$-band (2--4\,GHz) and with LOw Frequency ARray (LOFAR) at 144\,MHz.}
   {We measure vertical intensity profiles and exponential scale heights. We re-image both JVLA and LOFAR data at matched angular resolution in order to measure radio spectral indices between 144\,MHz and 3\,GHz. Confusing point-like sources were subtracted prior to imaging. Intensity profiles are then fitted with cosmic-ray electron advection models, where we use an isothermal wind model that is driven by a combination of pressure from the hot gas and cosmic rays.}
   {We discover a large-scale radio halo on one (northwestern) side of the galactic disc. The morphology is reminiscent of a bubble extending up to 20\,kpc away from the disc. We find spectral ageing in the bubble which allows us to measure advection speeds of the cosmic-ray electrons accelerating from 300 to $600\,\rm km\,s^{-1}$. Assuming energy equipartition between the cosmic rays and the magnetic field, we estimate the bubble can be inflated by a modest 10\,\% of the kinetic energy injected by supernovae over its dynamical time-scale of 35\,Myr. While no active galactic nucleus (AGN) has been detected, such activity in the recent past cannot be ruled out.}
   {Non-thermal bubbles with sizes of tens of kiloparsec may be a ubiquitous feature of star-forming galaxies showing the influence of feedback. To determine possible contributions by AGN feedback, will require deeper observations.}

   \keywords{cosmic rays -- galaxies: magnetic fields -- galaxies: fundamental parameters -- galaxies: star formation -- radio continuum: galaxies}


\titlerunning{A 20-kpc radio bubble in NGC\,4217}
\authorrunning{V.~Heesen et al.}

   \maketitle
%

\section{Introduction}

Galaxies in the radio continuum observed at gigahertz frequencies show the presence of cosmic-ray electrons and star formation. Cosmic-ray electrons are related to the presence of massive stars that end their lives as supernovae thereby accelerating these electrons to relativistic energies. Because the cosmic-ray electrons can travel, the resulting radio continuum image looks like a  smeared out version of the star-formation distribution \citep{murphy_08a,vollmer_20a}. This becomes particularly noticeable in the case of edge-on galaxies, where we see the galaxies from the side. Instead of a thin disc we see indeed a thick radio disc with a scale height of $\sim$1\,kpc, far in excess of the 100\,pc or so gaseous scale height of thin star-forming disc \citep{heesen_09a,heesen_16a,heesen_18a,krause_18a}. This extra-planar emission is referred to as `radio halo' and not to be confused with similarly named structures in clusters of galaxies. What makes the study of radio haloes so worthwhile is that it allows us a unique opportunity to study the transport of cosmic-ray electrons and the structure of magnetic fields. Both are important to understand galaxy evolution and, yet, are difficult to study both in simulations and in observations \citep[see][for reviews]{heesen_21a,ruszkowski_23a}.

In order to improve the situation on the observational side, the Continuum HAloes in Nearby Galaxies - an EVLA Survey project was initiated \citep[CHANG-ES;][]{irwin_12a,irwin_24a}. CHANG-ES is a radio continuum survey of 35 nearby galaxies including $L$- (1--2\,GHz) and $C$-band (4--6\,GHz) data with full polarisation. This project has lead to the realisation that radio haloes in star-forming galaxies are 
ubiquitous. In the 35-galaxy CHANG-ES sample, at least 17 galaxies show thick radio discs with scale heights of order 1\,kpc \citep{krause_18a, miskolczi_19a, mora_19a, schmidt_19a, stein_19a,stein_20a,heald_22a,stein_23a}. The remaining galaxies are either merging or have strong nuclear activity, so that the origin of a radio halo cannot be clearly distinguished. Nevertheless, it is becoming clear now that for star-forming galaxies a radio halo seems to be the norm at least for those cases where the star formation activity is sufficiently concentrated in the disc \citep{vasiliev_19a}.

Radio haloes show the presence of extra-planar cosmic-ray electrons (CRE) and magnetic fields. The CRE diffuse either out of the thin gaseous disc (scale heights order of 100\,pc) into the halo; alternatively, they are advected in a galactic wind. In addition, cosmic rays can stream along magnetic field lines at the Alfv\'en speed \citep{zweibel_13a}. Realistic models for galactic winds, should ideally take all these effects into account \citep{yu_20a}. The shape of radio haloes is sometimes rather boxy \citep{heald_22a}, with the halo extending vertically above and below the star-forming mid-plane. Their structure is clearly different from that of conical outflows from nuclear starbursts and bubble-like features. In fact, radio haloes  usually show no edge-brightening at all in total power radio continuum. In contrast, starburst galaxies often have biconical, limb brightened outflows visible in H\,$\alpha$ emission \citep{veilleux_05a}. They may be difficult to distinguish from the kiloparsec-scale radio-emitting outflows that approximately half of Seyfert-type galaxies show \citep{gallimore_06a}. The galaxy NGC\,6764 has both an active galactic nucleus (AGN) and a circumnuclear starburst; it is possibly the best example for limb-brightened radio bubbles driven by a combination of both feedback mechanisms  \citep{hota_06a}. AGN feedback is indistinguishable from that by star formation if neither jet-, lobe-, or bubble-like features are seen in radio continuum images \citep[see also][]{condon_92a}.

Some of the most notable structures in radio haloes are large scale-magnetic fields. Radio polarisation studies at centimetre wavelengths show that the large-scale magnetic field is parallel to the galactic planes in inner discs, but rising up from the mid-plane and outwards in the outer discs suggesting an X-shape structure in the halo \citep{krause_09a,heesen_09b, soida_11a, stein_20a, krause_20a}. Such a magnetic structure should also be traced in the total power radio continuum emission as well. Surprisingly though, no mirroring structure has been detected directly in total intensity maps at high-frequencies where most polarisation studies have been accomplished. Possibly because high-frequency observations trace only higher electron energies, only younger CRE in discs and star-forming regions are observed. However, magnetic fields in the haloes can  best be detected in synchrotron emission from old CRE, emitting at low-radio frequencies. With new $S$-band (2--4\,GHz) data we are enriching the CHANG-ES survey using deep polarimetric C-configuration observations aiming to cover a broad range of galaxy properties and thus strengthen our overall analysis. Polarised emission in $S$-band is the best probe for magnetic fields in the galaxies' faint halos because the frequency is at the transition between Faraday-thin and -thick regimes; also, spectral ageing is less severe than at $C$-band. Recent promising results obtained in $S$-band in nearby galaxies comprise NGC\,628 \citep{mulcahy_17a} and M\,51 \citep{kierdorf_20a}.

In this work we present a first look at the CHANG-ES $S$-band data using NGC\,4217 as an example. This galaxy was studied previously in detail by \citet{stein_20a} who used radio continuum data in $L$- (1--2\,GHz) and $C$-band (4--6\,GHz) as well as low-frequency data from the LOw Frequency ARray \citep[LOFAR;][]{vanHaarlem_13a} at 144\,MHz. They found strong hints of outflows: an active disc--halo interface with many bubbles of a few hundred parsec size and X-shaped magnetic fields in the halo. This galaxy was also studied by \citet{alton_00a} who searched for galactic winds and found `dust chimneys' in an optical absorption study \citep[see also][]{howk_99a}. Yet observations at other wavelengths do not find any indications for outflows along the minor axis. There is no extra-planar neutral hydrogen detected in the form of H\,{\sc i} emission \citep{zheng_22a}. Also, H\,$\alpha$ emission is only found in the form of two faint plumes on the south-western side of the disc indicating the presence of diffuse ionised gas \citep[][see also \citet{stein_20a} who find a plume on the north-eastern as well]{rand_96a}. \citet{hodges-kluck_16a} find extra-planar ultraviolet emission with {\it GALEX}; but this emission is not associated with outflows or extra-planar ionised gas, hence likely a reflection nebula created by dust. Similarly, X-ray observations of NGC\,4217 show no significant extra-planar diffuse X-ray emission from hot gas \citep{li_13a}.

In this paper, we will present the discovery of a new large-scale radio halo that has so far evaded detection. This galaxy shows no present AGN activity, neither as a nuclear source \citep{irwin_19a} or any associated variability \citep[cf.][where such a case is found in the CHANG-ES galaxy NGC\,4845]{irwin_15a}. Nonetheless, the presence of AGN activity cannot be ruled out conclusively, particularly not in the recent past. In fact there are several examples of galaxies discussed in the literature which possess radio bubbles such as in NGC\,6764 \citep{hota_06a}, NGC\,3079 \citep{duric_83a}, and the Circinus galaxy \citep{elmouttie_98a} These bubbles range in size from sub-kiloparsec to tens of kiloparsec \citep[see also][]{gallimore_06a}, and find that all of them have AGN at the centre. NGC\,4217 is potentially a unique case where a minor-axis radio bubble has been detected but it does not show any sign of AGN. However, given this rarity and considering the well-known episodic nature of jetted Seyfert galaxies, it is at least possible that the bubble is a relic from an AGN jet activity rather than created by galactic wind from star formation. 

This paper is organised as follows: Section\,\ref{s:data} presents our observations and data reduction. In Sect.\,\ref{s:results} we present our results with the main discovery of a new large-scale radio bubble. In Sect.\,\ref{s:discussion} we discuss the nature and origin of the bubble. We conclude in Sect.\,\ref{s:conclusions}. We assumed a distance to NGC\,4217 of $20.6$\,Mpc \citep{wiegert_15a}.

\section{Data}
\label{s:data}

\subsection{Observations and data reduction}
\label{s:observations}

Observations were taken with the {\it Jansky} Very Large Array (JVLA) in the $S$-band in the frequency range of 2--4\,GHz using full polarisation. We used C-configuration which provides us with a nominal resolution of around $7\arcsec$ using Brigg's robust weighting. The field of view as given by the diameter of the primary beam which is $15\arcmin$. The largest angular scale that we can image is $8\farcm 2$. Observations were taken in standard fashion with 2048 channels of 100\,kHz bandwidth each. 

\begin{figure*}
    \centering
    \includegraphics[width=\linewidth]{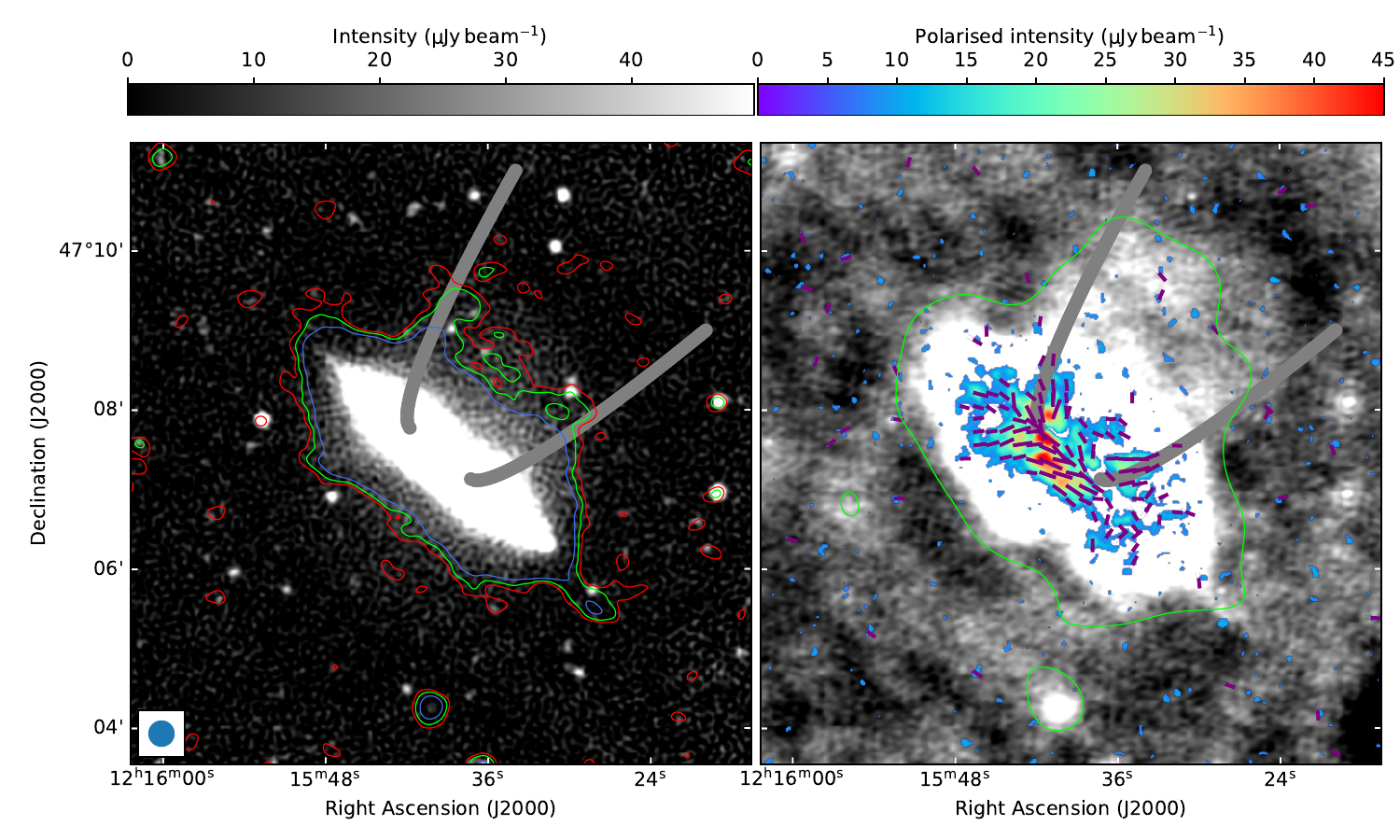}
    \caption{Radio continuum emission in the halo of NGC\,4217. {\it Left:} The grey-scale background map shows the JVLA 3\,GHz radio continuum emission at $7\arcsec$ angular resolution. Contour lines show the 2, 3, and 5$\sigma$ levels at 144\,MHz (red, green, and blue) observed with LOFAR at $20\arcsec$ angular resolution (indicated by the filled circle in the bottom left corner). Confusing background sources were subtracted in the $20\arcsec$ map. {\it Right:} LOFAR radio continuum emission at 144\,MHz imaged with enhanced diffuse extended emission at $20\arcsec$ angular resolution (for details see Sect.\,\ref{ss:radio_spectal_indices}). The contour line is at 320\,\uint with a resolution of $40\arcsec$, equating to $0.8\sigma$. Confusing background sources were subtracted. Coloured emission displays polarised intensity at $\approx$$9\arcsec$ resolution at 3\,GHz observed with JVLA and corresponding vectors show the orientation of the magnetic field (the vector length is proportional to the polarised intensity). Thick grey lines show the assumed shape of the CRE transport in the superbubble (Sect.\,\ref{ss:spix}).} 
    \label{fig:n4217}
\end{figure*}

\begin{figure}
    \centering
    \includegraphics[width=\linewidth]{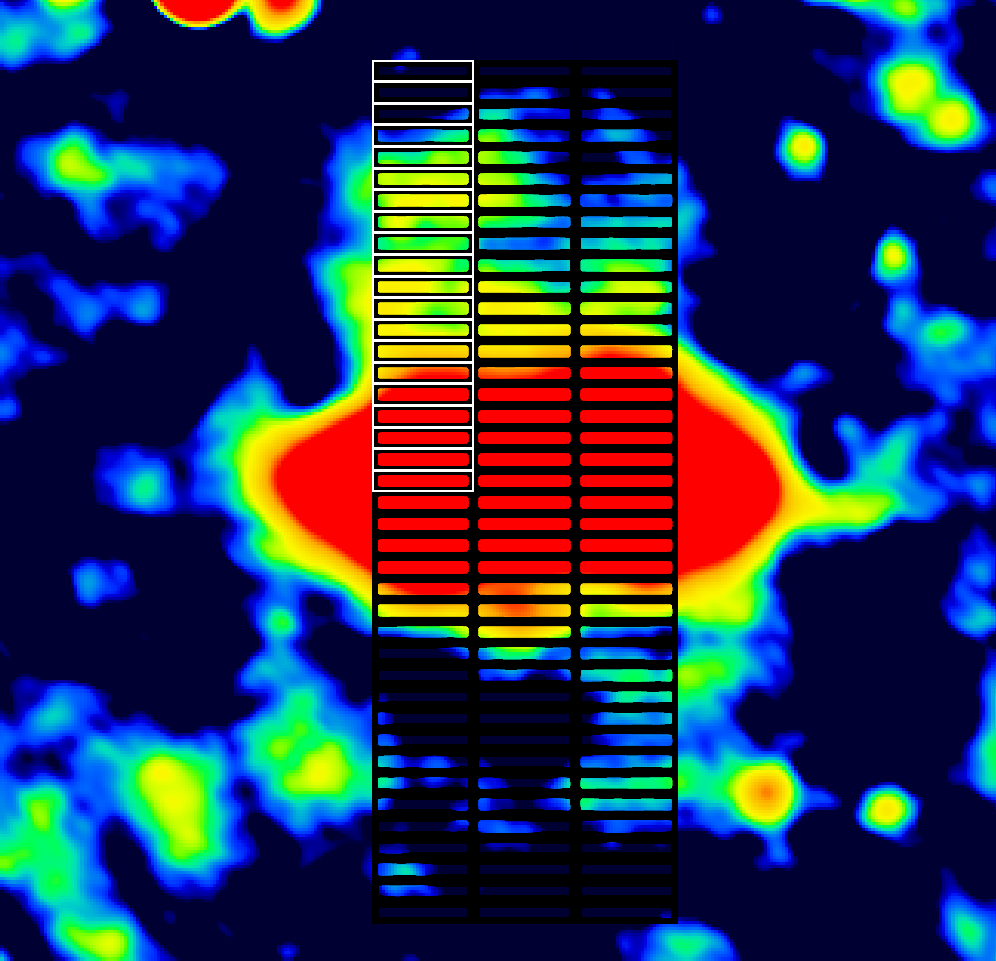}
    \caption{Position of vertical intensity profile of the superbubble in NGC\,4217 overlaid on the LOFAR map with boosted extended emission. We use the top-left strip for the analysis (additionally highlighted in white).  We are probing the potential shell on the northeastern side of the superbubble (cf.\, Fig.\,\ref{fig:n4217}, right panel).}
    \label{fig:strips}
\end{figure}

\begin{figure}
    \centering
    \includegraphics[width=\linewidth]{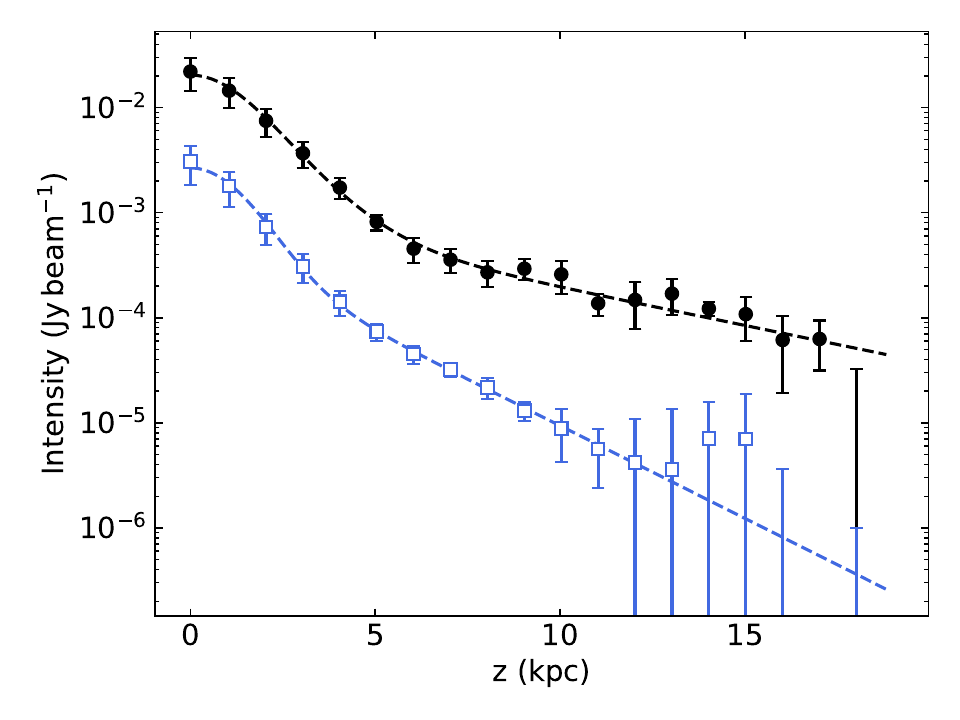}
    \caption{Vertical intensity profiles of the superbubble in NGC\,4217. Black filled data points show 144\,MHz emission and blue unfilled data points show 3\,GHz emission. Dashed lines show two-component exponential fits accounting for the limited angular resolution.}
    \label{fig:profile}
\end{figure}

\begin{figure}
    \centering
    \includegraphics[width=\linewidth]{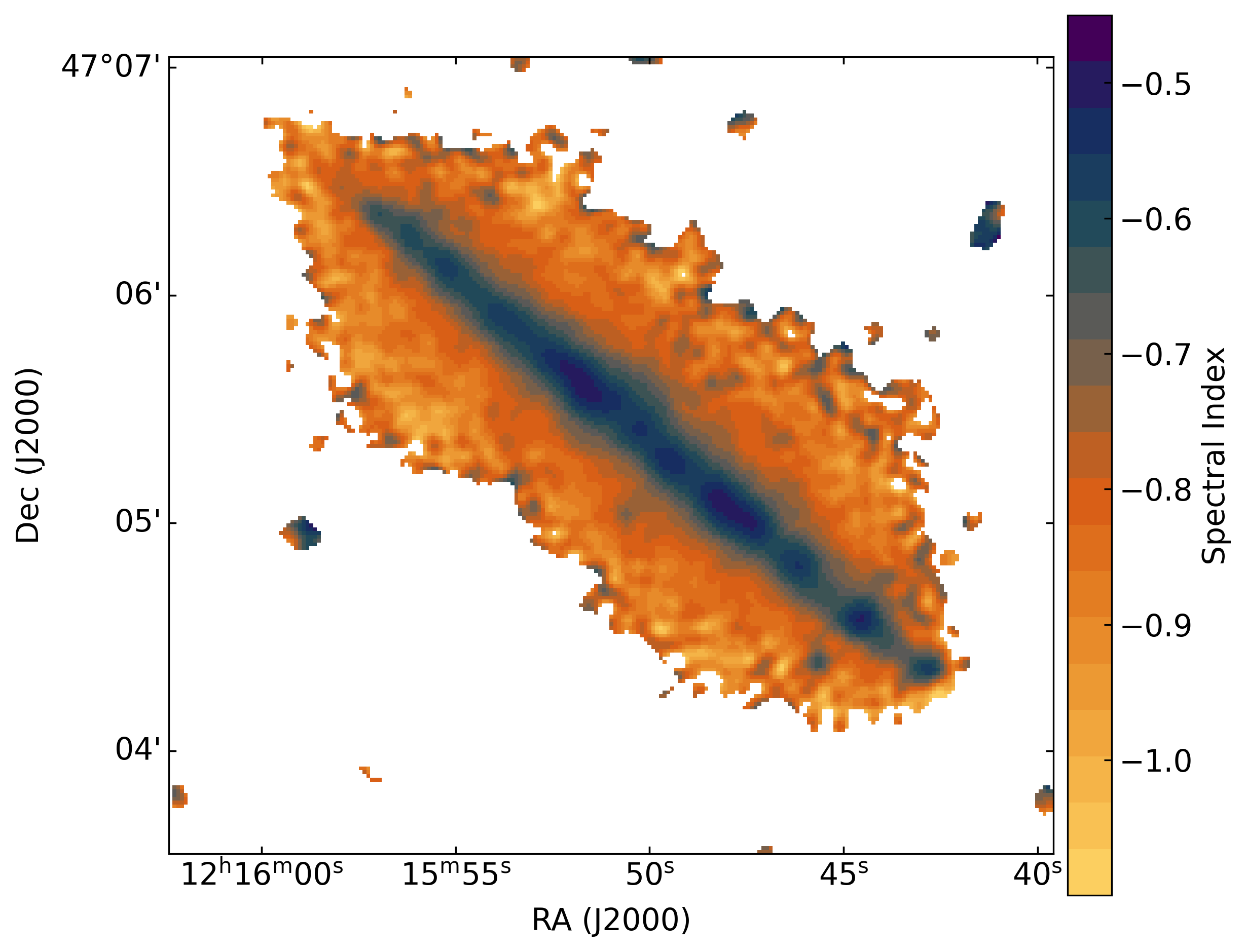}
    \caption{Radio spectral index between 144 and 3000\,MHz as false-colour representation at $7\arcsec$ angular resolution. Colour-code adopted from \citet{english_24a}.}
    \label{fig:spix}
\end{figure}

Our data are reduced with the Common Astronomy Software Applications \citep[{\sc casa};][]{casa_22a} version {\sc pipeline} $6.4.12$. For Stokes $I$ we used the VLA calibration pipeline without any further calibration or data flagging except for self-calibration. The data were imaged with {\sc wsclean} version $2.9$ using a multi-scale {\sc clean} algorithm \citep{offringa_14a}. The $S$-band data have a field of view of $15\arcmin$, so that we have made maps with a size of approximately $1.8$ times the primary beam extent. We then performed one self-calibration in phase only before creating final maps. We created one map with $7\arcsec$ angular resolution using $\mathtt {robust} = 0.5$ and one with $20\arcsec$ angular resolution using $\mathtt {robust} = 1.5$ and an additional Gaussian taper of $10\arcsec$ width. Final maps were restored with Gaussian {\sc clean} beams using $7\arcsec$ and $20\arcsec$ full width at half maximum (FWHM) where the size is closely matched to the native angular resolution. We also applied the primary beam correction with {\sc wsclean}. The rms map noise values are $3.5$ and $5$\,\uint\ at $7\arcsec$ and $20\arcsec$ resolution, respectively. This is only 10\,\% higher than the thermal noise level as estimated from the VLA exposure calculator with consideration of the flagged data fraction (30\,\%).

\subsection{Radio spectral indices and polarisation}
\label{ss:radio_spectal_indices}

In order to measure radio spectral indices, we also re-imaged the LOFAR 144\,MHz data at identical angular resolutions with {\sc wsclean}. These data were already presented in \citet{heesen_22a} and derive from the LOFAR Two-metre Sky Survey data release 2 \citep[LoTSS-DR2;][]{shimwell_22a}. Radio spectral indices were then calculated in the usual fashion with the corresponding uncertainties \citep[e.g.][]{heesen_22a}. In order to avoid the influence of point-like background sources, we subtracted them before making the low resolution maps. This was done by fitting Gaussian functions to the high-resolution maps using {\sc pybdsf} \citep{mohan_15a} and then subtracting them from the $(u,v)$ data. 
We also created another low-resolution LOFAR map using $\mathtt{robust} = 0.35$ in order to boost specifically extended emission, again at an angular resolution of $20\arcsec$; this map has a rms noise of $100$\,\uint\, which is reasonably good for LOFAR at this frequency and resolution \citep{heesen_22a}. We used the LOFAR low-resolution map only for illustrative purposes in order to show the weak extended halo emission. The reason is that while the sensitivity to diffuse emission is high, the elevated side-lobe levels for positive values of the $\mathtt{robust}$ parameter make a reliable deconvolution of LOFAR data with the {\sc clean} algorithm very hard \citep{sridhar_18a}. We subtracted the thermal contribution using a combination of H\,$\alpha$ and mid-infrared emission as described in \citet{stein_23a}, which is based on the prescription by \citet{vargas_18a}. The thermal contribution from free--free emission is estimated to be less than 20\,\% at 3\,GHz within $\approx$2\,kpc from the mid-plane and less than 10\,\% in the halo \citep{stein_19a}. At 144\,MHz the thermal fraction is even lower.

For polarisation, we performed standard data calibration. In brief, we inserted polarisation models of our primary polarisation calibrator 3C\,286 and adjusted our polarisation angle accordingly \citep{perley_13a}. The instrumental polarisation was calibrated with the help of the unpolarised calibrator J$1407+2827$. The $(u,v)$ data were then imaged with a Brigg's robust weighting of $\mathtt{robust} = 1.5$ resulting in a synthesised beam size of $8\farcs 52\times 7\farcs 89$ (position angle of $165\fdg 1$). We did not use any self-calibration as this was not necessary with the rms noise values $2.3$\,\uint\ in the Stokes $Q$ and $U$ maps. These maps are then converted into linear polarised intensity, correcting for positive Ricean bias as described in \citet{wardle_74a}. We also calculated the polarisation angle of the $E$-vectors. Galactic rotation measure was estimated as $8.1\pm 4.0\,\rm rad\,m^{-2}$ using the 
model from \citet{hutschenreuter_20a}. Polarisation angles were corrected accordingly subtracting a constant angle of $4\fdg 6$, equivalent to the rotation at a wavelength of 10\,cm. In order to obtain the orientation of the magnetic field, we rotated the polarisation angle of the $E$-vectors by $90\degr$. These angles are not corrected for internal Faraday rotation.

\section{Results}
\label{s:results}

\subsection{Morphology of the radio halo}
\label{ss:radio_halo}

The total power radio continuum distribution is presented in Fig.\,\ref{fig:n4217}. We found that already the high-resolution $7\arcsec$ map showed a distinctive extension in the north-west at 3\,GHz. However, this component is very diffuse and hence shows better in the low resolution maps. The contour lines shows that the distribution at $20\arcsec$ angular resolution at 144\,MHz shows a horn-like structure but is also extended on the northwestern halo. 
This gives us confidence we have detected a real structure. 
Using the LOFAR $20\arcsec$ map with the short baselines boosted we detect this structure even further out to approximately $200\arcsec$ distance from the galactic mid-plane, equivalent to a projected distance of 20\,kpc (Fig.\,\ref{fig:n4217}, right panel). 

The morphology of the radio continuum emission at 144\,MHz (Fig.\,\ref{fig:n4217}, right panel) is reminiscent of a bubble. The emission is boosted along the bubble walls with a slight depression in the centre of the bubble. The northeastern edge of the bubble is in particular prominent and image seems to indicate the potential presence of a shell on this side. This emission and the horn-like structure suggests that there is some sort of real edge on the northeastern side of the bubble.

\subsection{Vertical radio intensity profiles}
\label{ss:profiles}

We constructed vertical intensity profiles as described in \citet{stein_23a}. For this we used the {\sc boxmodels} of {\sc nod3} \citep{mueller_17a}. We chose three strips that approximately cover the width of the bubble with a strip width of $50\arcsec$ each, sampling the intensity every $10\arcsec$ distance from the mid-plane. The position of the strips is shown in Fig.\,\ref{fig:strips}. We restricted the analysis to the northeastern side of the superbubble as the intensity measurements have the highest signal-to-noise ratio (highlighted in white in Fig.\,\ref{fig:strips}).

The vertical intensity profiles shown in Fig.\,\ref{fig:profile} 
reveal a new, distinct
halo component, the purported radio superbubble. The break in the intensity profiles occurs at approximately 5\,kpc where the slope of the profiles flatten significantly. The scale heights of the bubble component are $5.9\pm 1.1$ and $2.9\pm 0.3$\,kpc, at 144 \,MHz and 3\,GHz, respectively. These scale heights are a factor of a few larger than the typical scale heights of $\approx$$1$\,kpc in edge-on galaxies  \citep{krause_18a}.

\subsection{Magnetic fields}
\label{ss:magnetic_fields}

NGC\,4217 presents a prominent example of the so-called X-shaped magnetic fields that are sometimes seen in edge-on galaxies \citep{krause_20a}. This was already found with data at 5\,GHz \citep{stein_20a} and is confirmed with the new $S$-band data as shown in Fig.\,\ref{fig:n4217} (right panel). The polarised emission and the orientation of the magnetic fields partially align with the purported base of the radio bubble. Such an alignment may be caused if the magnetic fields are concentrated in the walls of the bubble. The magnetic field structure is in part also reminiscent of the the polarisation of the {\it Fermi} bubbles in the Milky Way observed in $S$-band \citep{carretti_13a}. Analytical models as well as simulations of galactic winds show that field lines wind up in a helical fashion with the field lines anchored in the disc \citep{ptuskin_97a,thomas_23a}.

\citet{stein_20a} estimated the magnetic field strength using energy equipartition with the revised equipartition formula of \citet{beck_05a}. The central region of the galaxy shows total magnetic field strengths of $11.0\,$\umag; the mean disc field strength is $9.0$\,\umag; and the halo magnetic field is roughly constant at a value of $7.4$\,\umag. These values are in good agreement with the values of other spiral galaxies \citep{beck_19a}.

\begin{figure}
    \centering
    \includegraphics[width=\linewidth]{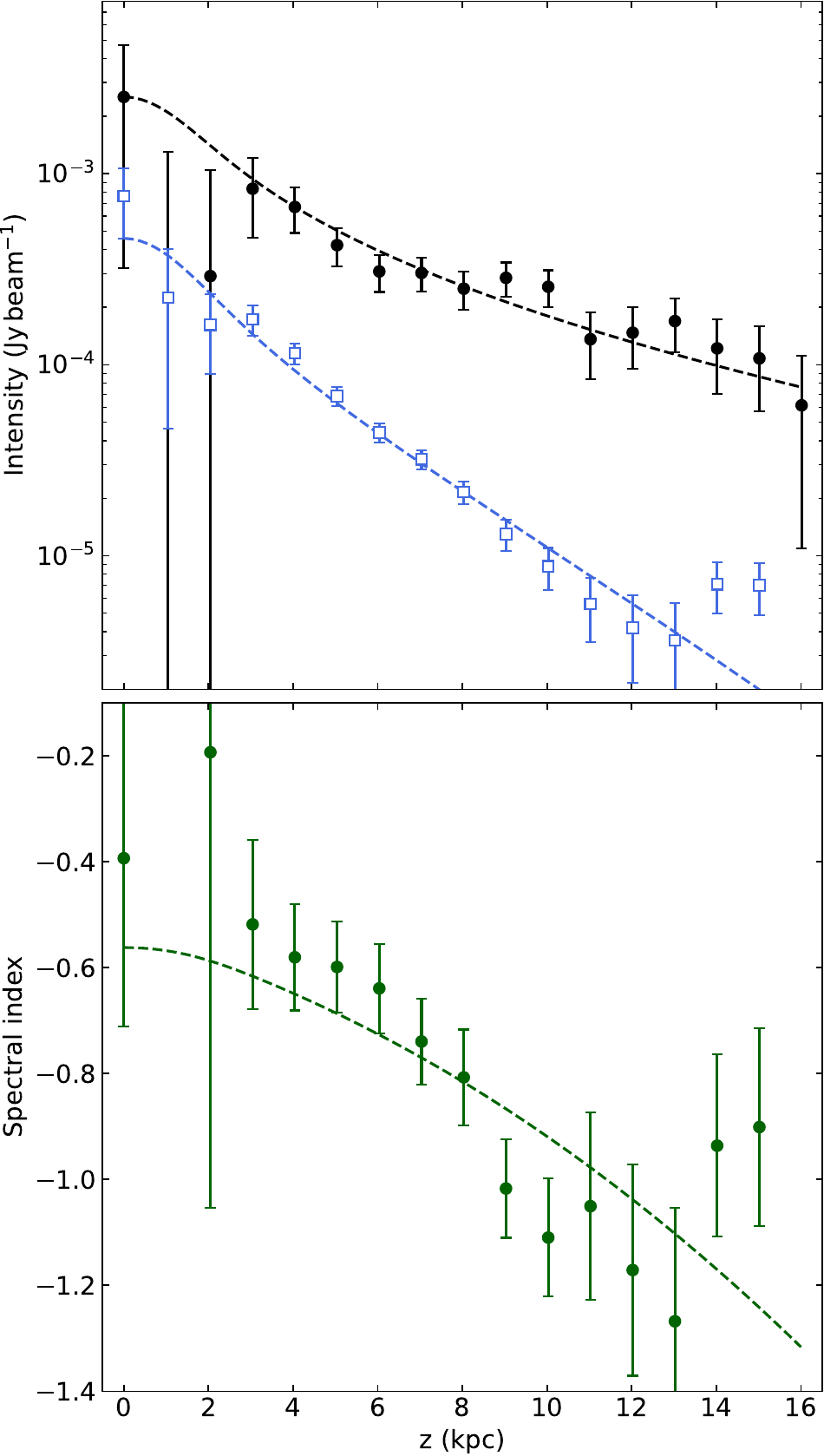}
    \caption{Radio continuum emission and radio spectral index in the superbubble. The thick disc component was subtracted. We present the LOFAR 144\,MHz (top graph, black) and $S$-band 3\,GHz (top graph, blue) intensity profiles as well as the corresponding radio spectral index profile (bottom graph). Best-fitting {\sc spinnaker} wind models are shown as a dashed lines.}
    \label{fig:spi}
\end{figure}

\begin{table}
\caption{Best-fitting wind solution for the superbubble.}
\label{table:wind}
\centering
\begin{tabular}{l c}
\hline
Parameter & wind  \\
\hline
$B_0$ ($\upmu\rm G$)$^a$ & $11.0$ \\
$v_{\rm c}$ ($\rm km\,s^{-1}$)$^{b}$ & $280^{+110}_{-60}$ \\
$z_0$ (kpc)$^c$ & $5.7^{+1.8}_{-2.6}$ \\
$\beta$$^d$ & $1.2^{+0.2}_{-0.2}$ \\
$\gamma$$^e$ & $2.1^{+0.15}_{-0.10}$ \\
$\chi_\nu^2$$^f$ & $1.2$ \\
\hline
\end{tabular}\\
\flushleft
\textbf{Notes.} \\
(a) Total magnetic field strength in the disk (fixed);\\
(b) Wind speed at the critical point; \\
(c) Scale height of the flux tube (see Eq.~\ref{eq:flux_tube});  \\
(d) Power-law index for the flux tube (see Eq.~\ref{eq:flux_tube}); \\
(e) CRE injection spectral index; \\
(f) Reduced $\chi^2$. \\
\end{table}

\begin{figure}
    \centering
    \includegraphics[width=\linewidth]{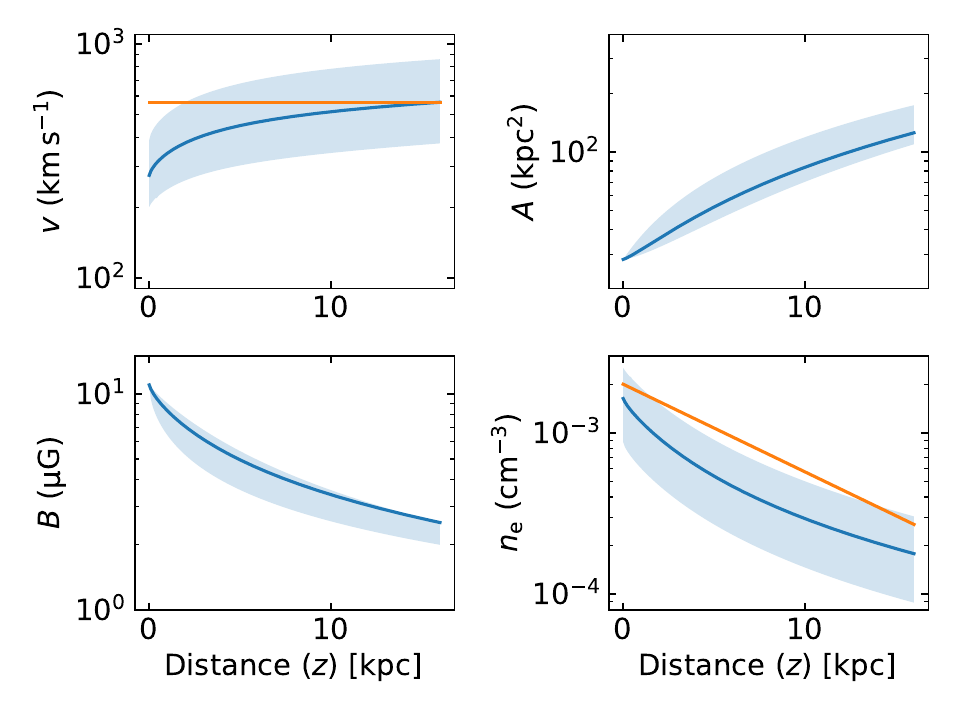}
    \caption{Wind model in NGC\,4217. We show the vertical profiles of the relevant parameters: wind velocity (top left), cross-section of the flux tube (top right), magnetic field strength (bottom left), and thermal electron density (bottom right). The estimate for the escape velocity is shown as solid line (top left panel). In bottom right panel, we also show a possible profile of the electron density from X-ray data using a scale height of 8\,kpc with arbitrary scaling.}
    \label{fig:vel}
\end{figure}

\subsection{Cosmic-ray electron transport}
\label{ss:spix}

In order to study the transport of CRE in the superbubble, we used the vertical radio spectral index profile between 144\,MHz and 3\,GHz. As the superbubble shows up as distinct component in the intensity profiles with a large scale height (Sect.\,\ref{ss:profiles}), we first subtracted the thick radio disc. The resulting intensity profiles in the radio bubble together with the spectral index profile are shown in Fig.\,\ref{fig:spi}. We can see that the radio spectral index gradually steepens between the disc and edge of the halo. The outer edge is here shown at 16\,kpc distance where the last significant detection is made at 3\,GHz. The gradual, almost linear, steepening can be well fitted with a cosmic-ray advection model for the electrons implemented in {\sc spinnaker} \citep[SPectral INdex Analysis of K(c)osmic-ray Electron Radio-emission;][]{heesen_16a,heesen_18a}. We used the flux tube model  described in \citet{heald_22a} that assumes a helical magnetic field in the outflow \citep[see also][]{stein_23a}. 

The flow of the plasma is governed by the Euler equation:
\begin{equation}
    \rho v \diff{v}{z} = -\diff{P}{z}  - g\rho, 
    \label{eq:euler}
\end{equation}
where $P=P_{\rm gas}+P_{\rm CR}$ is the combined gas and CR pressure, $g$ is the gravitational acceleration, and $\rho$ the gas density. Here we assume $P_{\rm CR}=1/3u_{\rm CR}$ and $P_{\rm gas}=2/3u_{\rm gas}$, where $u_{\rm CR}$ and $u_{\rm gas}$ are the energy densities of the cosmic-ray and thermal gas, respectively. We assume $u_{\rm CR}=4.8\times 10^{-12}\,\rm erg\,cm^{-3}$ (in equipartition with the magnetic field in the disc at $z=0$, so $u_{\rm CR}=u_{\rm B}$ with $u_{\rm B}=B_0^2/(8\uppi)$ with $B_0=11\,\rm \upmu G$) and $u_{\rm gas}=1.2\times 10^{-12}\,\rm erg\,cm^{-3}$. \citet{li_13a} studied the thermal gas energy density in galactic haloes and found that it is equivalent to a hot gas with a temperature of $0.5\,\rm keV$ and a number density of $2\times 10^{-3}\,\rm cm^{-3}$.  The same authors found in their 70\,ks {\it Chandra} observation of NGC\,4217, that the total $0.5$--$2$\,keV luminosity from hot gas is $2\times 10^{39}\,\rm erg\,s^{-1}$ and the vertical scale height is $\approx$4\,kpc. This suggests an electron scale height of approximately 8\,kpc.

We assume the following functional term for the cross-sectional area (parallel to the disc):
\begin{equation}
    A(z) = A_0 \left [1 + \left (\frac{z}{z_0}\right )^\beta \right ],
    \label{eq:flux_tube}
\end{equation}
where $z$ is the distance to the disc (assuming cylindrical symmetry). This form describes an expanding flow, which has been used previously in semi-analytic 1D wind models \citep[e.g][]{breitschwerdt_91a}. Here $A_0=\uppi r_0^2$ is the area in the mid-plane and $z_0$ is the flux tube scale height. The parameter $\beta$ describes the shape of the flux tube, where $\beta=2$ means the flux tube has a constant opening angle, so that its shape can be described as a conical frustum. We assume an initial outflow radius of $r_0=2\,\rm kpc$.

We also require an equation that governs the magnetic field strength:
\begin{equation}
    B = B_0 \left (\frac{r}{r_0}\right )^{-1} \left ( \frac{v}{v_0}\right )^{-1}, 
\end{equation}
where $B_0$ is the magnetic field strength in the galactic midplane, and $r_0$ and $v_0$ are the midplane flow radius and advection speed, respectively. This is the expected behaviour for radial and toroidal magnetic field components in an axisymmetric, accelerating, quasi-1D flow \citep{baum_97a}.

Then we can write the Euler equation in the following way:
\begin{equation}
    \rho v \diff{v}{z} = -v_{\rm c}^2 \diff{\rho}{z} -g\rho.
\end{equation}
This equation contains only the velocity $v$ and the density $\rho$. The density can be eliminated with the continuity equation $va\rho=\rm const.$ and using the definition of the composite sound speed $v_{\rm c}^2=P/\rho$. The wind velocity is equivalent to the composite sound speed in the critical point. CRE are injected in the disc plane with a power-law in number density  $N\propto E^{-\gamma}$ as function of CRE energy $E$, where $\gamma$ is the injection spectral index. Spectral ageing of the CRE is then created by a combination of synchrotron and inverse Compton losses. 

The best-fitting advection model is shown in Fig.\,\ref{fig:spi} as solid lines with the best-fitting parameters summarised in Table\,\ref{table:wind}. We find that the wind speed rises from approximately $300$ to $600\,\rm km\,s^{-1}$ at the edge of the bubble. This is approximately equivalent to the escape velocity $v_{\rm esc}\approx 3v_{\rm rot}$ \citep{veilleux_20a}, where $v_{\rm rot}=188\,\rm km\,s^{-1}$ is the rotation speed \citep{heesen_22a}. The best-fitting parameters are presented in Table\,\ref{table:wind}. In Fig.\,\ref{fig:vel} we show vertical profiles of the relevant physical parameter of the wind. The advection speed within the bubble implies a dynamical time-scale of 35\,Myr if the bubble expands as fast as the cosmic-ray electrons. We find tentative evidence of cosmic-ray re-acceleration at approximately 14\,kpc distance from the disc, visible as spectral flattening. This would mean our advection time-scale is a lower limit. The magnetic field strength decreases from $11\,\rm \upmu G$ in the mid-plane to $\approx$3\,\umag\ at the edge of the bubble.

\begin{figure}
    \centering
    \includegraphics[width=\linewidth]{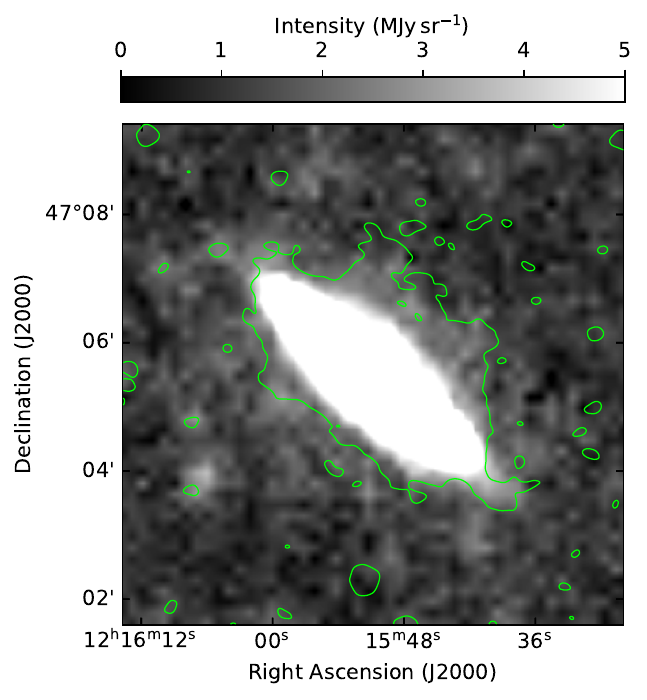}
    \caption{Overlay of radio continuum contours on far-infrared emission. Contours show 144\,MHz emission and are at $2\,\sigma$ equivalent to 200\,\uint (compare with Fig.\,\ref{fig:n4217}, left panel). Grey-scale image shows {\it Herschel}  $350\,\rm \upmu m$ emission.}
    \label{fig:dust}
\end{figure}

\section{Discussion}
\label{s:discussion}

\subsection{The nature of the radio bubble}
\label{ss:bubble}

The apparent starburst-driven radio bubble on multiple kiloparsec scales we have discovered is, to our knowledge, a unique feature in an external, star-forming, edge-on galaxies. However, we should keep in mind that such structures may occur in other galaxies too, albeit hidden in the general radio halo. 
If the bubble would be a factor of only two-to-three smaller it would be hard to distinguish from the radio halo. Across a (5\,kpc)$^2$ region we measure a spectral luminosity of $\sim$$10^{20}\,\rm W\,Hz^{-1}$ which is about 1\,\% of the typical 144\,MHz luminosity of a star-forming galaxy \citep{smith_21a}. Hence, these bubbles are rather difficult to detect. There are, however, a few other galaxies with an extended third component with scale heights of a few kiloparsec, for example in NGC\,4631 \citep[][and M.\,Stein, private communication]{stein_23a}. Of course there are galaxies that have extended lobes in their haloes, such as M\,106 \citep[see][]{zeng_23a} and NGC\,3079 \citep[e.g.][]{veilleux_94a}, but these may be due to a combination of star formation and AGN activity. NGC\,4217 does not currently  host an active AGN \citep{irwin_19a}. On the other hand, an AGN may be intermittent only and the lack of a present AGN in the galaxy alone cannot be used to conclude that the radio bubble is due to a starburst. In addition to the argument of the energetics, the morphology of the bubble may be more useful, which resembles the radio halo in NGC\,3079, which  probably contain the most well defined bubbles not necessarily produced by AGN.

Following the discovery of polarised emission from the {\it Fermi} bubbles in the centre of the Milky Way \citep{carretti_13a}, there arose the question whether such structures may ever be detected in external galaxies. We note that in the Milky Way there is no clear detection of radio continuum emission from the {\it Fermi} bubbles, possibly because of confusion with the foreground or background. However, it may be worth mentioning that there are some smaller scale analogues in the Galactic centre where such radio bubbles are discovered with MeerKAT \citep{Heywood2019a}. Therefore the external view on NGC\,4217 may be a unique opportunity to study radio continuum emission from such a feature. What is clear in NGC\,4217 is that both CRE and magnetic fields must be present inside the bubble in order for us to detect radio continuum emission.

The magnetic field structure we propose, helical fields, are found to be a common result of dynamo action in galaxies as summarised in \citet{henriksen_21a}. An application of the theory to NGC\,4631 is found in \citet{woodfinden_19a} where some evidence was found for helical magnetic fields. \citet{henriksen_21a} studied the separate effects of stellar turbulence, halo winds, and halo lags on the dynamo field. They found that it is difficult to get a helical field 
 that develops into a geometric X-field without winds. Halo lags accentuate the winding of the helix, which is likely to isotropise the synchrotron radiation. The field can be wound into flat helices near the disc.  An important feature of the theory is that it is scale invariant, at least between the stellar scale and the intergalactic scale. This means that it can be applied to kiloparsec-sized bubbles, taking the axis to be centred on the bubble. The same of magnetic fields should develop.
 
\subsection{Bubble energy and formation}
\label{ss:energy}

We may now compute how much energy is needed to inflate the bubble. 
If the bubble is assumed to be spherical with a radius of 10\,kpc the volume is $3.9\times 10^{67}\,\rm cm^3$. Expanding against a typical external pressure of $P_{\rm ext} \sim 1\,\rm eV\,cm^{-3}$, the energy to inflate the bubble can be estimated as 
$5/2 \ P_{\rm ext} V \approx $$1.6 \times 10^{56}\,\rm erg$. 
Now we can check whether the star formation activity is enough to inflate the bubble over its dynamical time-scale of 35\,Myr estimated from the advection time; This is a lower limit as the propagation speed of the bubble may be lower than the CRE transport speed. With a star-formation rate of $4.61\,\rm M_\sun\,yr^{-1}$ we expect a core-collapse supernova rate of $0.053\,\rm yr^{-1}$ \citep{murphy_11a}. With a canonical $10^{51}\,\rm erg$ of kinetic energy per supernova injected, the energy injection rate is $1.7\times 10^{42}\,\rm erg\,s^{-1}$. Hence, the total energy injected over the course of 35\,Myr is $1.8\times 10^{57}\,\rm erg$, which would be easily sufficient to inflate the bubble. Of course, not all the kinetic energy can be used to inflate the bubble as large fraction may be radiated away.

What might help to inflate the bubbles are cosmic rays. As 10\,\% of the kinetic energy injected by supernovae goes into the acceleration of cosmic rays \citep[a conservative estimate, see][]{grenier_15a}, this energy alone would suffice. Again, one might argue that not all cosmic-ray energy is funnelled into the bubbles, as the galaxy has also a normal radio halo. But what seems to be clear is that the energy from star formation is easily able to inflate the bubble and no AGN is needed. The average energy density of cosmic rays in the bubble is 15\,\% of the value in the disc, equating to $0.15B_0^2/(8\pi)$ or $6\times 10^{-13}\,\rm erg\,cm^{-3}$. This includes the energy of the protons, which dominate. Hence, the cosmic rays are able to provide a fair fraction (40\,\%) of the total energy that is needed in order to inflate the bubble.


One peculiarity is, of course, why did the bubble form only on one side of the galaxy? One hint as to why that might be comes from observations in X-rays and in the far-infrared. In X-ray emission  hot gas is only detected in the south-eastern halo (i.e., the one without a radio bubble), but not in the north-western halo \citep{stein_20a}. Although the X-ray data's signal-to-noise ratio is too low to map the intensities, the large-scale X-ray enhancements in the northwestern halo seem apparent, consistent with the large 95\,\% scale height. In contrast, there is extra-planar dust emission at $350\,\rm \upmu m$ as shown in Fig.\,\ref{fig:dust} only in the north-western halo. The far-infrared image is from the HERschel Observations of Edge-on Spirals \citep[HEROES;][]{verstappen_13a} project. Such an asymmetry may arise, for instance, if by happenstance there is a larger column of dense gas between the locus of star formation close to the midplane and the north-west hemisphere than between the star-formation and the south-east hemisphere.
In this case, star-formation heated plasma will escape down the steepest density gradient (into the south-east) while, in contrast, ion-neutral damping means that cosmic rays actually escape fastest via streaming \citep{kulsrud_69a,everett_11a} through the neutral gas on the north-west side.
Further, as thermally driven winds tend to be hotter than cosmic-ray driven winds \citep{girichidis_18a}, it seems consistent that dust grains can survive transport to large scale heights  in the putatively cooler, cosmic-ray-supported wind in the north-western halo while they are destroyed on the south-eastern side.
We note that, while this scenario seems self consistent, there do apparently need to be patches of sufficiently low column to allow the escape of sufficient ultraviolet emission to illuminate the dust grains on the north-west side \citep[cf.,][]{hodges-kluck_14a}.

\subsection{Radio haloes and galactic winds}
The shape of radio haloes is sometimes referred to as dumbbell-shaped, with the maximum extent not in the centre of galaxies but rather in the outskirts. A prominent example is NGC\,253 \citep{heesen_09a,kapinska_17a} and to a lesser extent NGC\,891 \citep{mulcahy_18a} as well as NGC\,4217 \citep{stein_20a}, and possibly a few other galaxies \citep{wiegert_15a}. One interpretation of such a structure was that the CRE suffer higher synchrotron losses in the centre of the galaxy and thus are not able to get out as far as in the galactic outskirts \citep{heesen_09a}. However, in light of our new finding of the superbubble, it appears that the `horns' of the dumbbell are indeed just the walls of the superbubble that are brighter due to limb brightening. This limb brightening requires a lateral density gradient of the magnetic fields and cosmic rays. Such a density gradient was suggested by early hydrodynamical simulations of starburst-driven superwinds \citep{heckman_00a}. They suggested that the coolest densest gas is associated with the swept-up shell of interstellar medium that propagates laterally in the plane of the galaxy. This process may be enhanced by entrainment and stripping of cool dense gas into the wind flowing out of the disk (through Kelvin--Helmholtz instabilities. Indeed early X-ray observations found such horns in NGC\,253 \citep{pietsch_00a} later to be shown part of a more extended structure \citep{bauer_07a}. What was so far missing is the connection with the radio halo, where such filled structures evaded detection until now. Hence, it appears likely that the horns of the dumbbell-shaped radio haloes are indeed the walls of superbubbles where hot thin gas is hard to detect in the radio continuum.

Alternatively, instead of a wind-driven bubble, we now consider AGN-driven outflows. This is motivated by observations where, starting from early examples of a few objects showing radio bubbles \citep{hummel_83a} to later a representative sample of ten objects with radio bubbles \citep{hota_06a}, the influence of AGN is likely at least. NGC\,4217 and possibly the Milky Way are  the only examples of radio bubbles without visible AGN feedback. Nevertheless we note, an AGN can quickly dominate feedback in galaxies once active. While the morphology of the radio continuum emission does not reveal a jet in NGC\,4217, this also not the case in other galaxies with radio bubbles that have an AGN (such as NGC\,3079, the Circinus galaxy, and NGC\,6764). Therefore the lack of any jet emission is not enough to conclusively rule out AGN feedback. More observations of radio bubbles in galaxies without AGN are needed in order to establish the case for purely wind-driven radio bubbles.

\section{Conclusions}
\label{s:conclusions}

We present new observations of the CHANG-ES galaxy NGC\,4217 in $S$-band (2--4\,GHz) which we combine with archival LoTSS-DR2 data at 144\,MHz.  This galaxy was known to have a radio halo extending to about 5\,kpc distance from the star-forming disc with a prominent X-shaped magnetic field in the halo \citep{stein_20a}. With the new, much more sensitive $S$-band data we detected a conspicuous extension of radio continuum emission in the north-western halo (Fig\,\ref{fig:n4217}, left panel). This prompted us to search again for extended emission in the LOFAR data previously studied. We found indeed a very extended faint component that had previously evaded detection. This component has the morphology of an edge-brightened bubble that extends out to a distance of 20\,kpc from the star-forming disc (Fig\,\ref{fig:n4217}, right panel). Such a structure is reminiscent of the {\it Fermi} bubbles in the Milky Way, although the size in NGC\,4217 is 2--3 larger in size (length) than the one in our own Galaxy.

Radio haloes reveal the presence of CRE and magnetic fields and thus have been suspected to be indicative of galactic winds and outflows \citep[e.g.][]{heesen_21a}. However, the morphology is mostly different from what galactic wind simulations predict, namely bubbles along the minor axis of the galaxy. In fact radio haloes have a boxy shape that extends over the star-forming disc \citep[e.g.][]{heald_22a}. With this discovery we show that at least in one case we do find bubbles of cosmic rays that can be explained by the star formation in the disc and thus may be good indicators of galactic winds. There are also other cases where there are hints of bubble-like structures both in the radio continuum as well as in other wavelengths. For instance, in the iconic starburst galaxy NGC\,253 prominent bipolar X-ray bubbles were observed \citep{pietsch_00a}. In NGC\,253, the radio continuum has a dumbbell-shaped radio halo with an X-shaped magnetic field tangentially aligned with the edge of the superbubbles in the halo \citep{heesen_09a,heesen_09b}. However, in contrast to NGC\,4217, in NGC\,253 there are no bubbles detected in total power radio continuum emission. As \citet{stein_20a} have shown NGC\,4217 is also a dumbbell-shaped galaxy with extended emission at 144\,MHz along the edges of the bubble (their fig.\,1). Yet another example with a possible bipolar outflow in the X-rays and X-shaped magnetic fields is NGC\,5775 \citep{heald_22a}. Although a well defined bubble is only detected on one side of the galactic disc, there are still signatures of bubble-like features on the other side, the magnetic field and the H\,$\alpha$ emission may trace the walls of the superbubble \citep[][their fig.\,8]{li_08a}.

In light of our new findings, we propose that the horns of dumbbell-shaped radio haloes are the edge-brightened boundaries of bipolar radio outflows. Cosmic rays are transported in these starburst-driven galactic winds,  streaming along X-shaped magnetic field lines \citep{krause_09a}. While the outflow in NGC\,4217 may be purely star-formation driven, the well defined outer boundary suggests that it is an episodic outflow event and the cosmic rays have not yet diffused away. This would favour an AGN-driven outflow, which agrees with the observed lack of gas in the halo. On the other hand, jet-driven bubbles often have random orientation to the disc \citep[e.g. in NGC\,3801;][]{hota_12a}, hindering the formation of bipolar outflows. We need deeper observations to discover a sample of purely starburst-driven radio bubbles. Thanks to the excellent sensitivity of LOFAR such discoveries are possible, showing for instance that galaxy-sized are quite common \citep{webster_21a}. Until then, AGN feedback will remain at least a distinct possibility which cannot be ruled out.

Summarising, the radio superbubble in NGC\,4217 reveals an ingredient in the observations of radio haloes that has been missing thus far. It is not clear why there are not more of such bubbles detected elsewhere, but clearly spectral ageing plays an important role. With future observations at 50\,MHz with LOFAR \citep{de_gasperin_21a} it may become possible to see more of these structures. What seems clear is that bipolar outflows of cosmic rays may be indeed commonplace in star-forming galaxies. They would fit nicely into what is predicted by theories of stellar feedback and may be one signature of cosmic ray-driven galactic winds \citep[e.g.][]{thomas_23a}. In the future, we may consider {\it XMM--Newton} follow-up observations searching for the extended hot halo of NGC\,4217, although it is not well detected in a 70\,ks {\it Chandra} observation \citep{li_13a}.

\begin{acknowledgement}
We thank the referee for a constructive report which helped to improve the paper. This work was performed in part at Aspen Center for Physics, which is supported by National Science Foundation grant PHY-2210452. We thank Rainer Beck for valuable comments on an early draft of the manuscript. This paper is based (in part) on data obtained with the International LOFAR Telescope (ILT). LOFAR \citep{vanHaarlem_13a} is the Low Frequency Array designed and constructed by ASTRON. It has observing, data processing, and data storage facilities in several countries, that are owned by various parties (each with their own funding sources), and that are collectively operated by the ILT foundation under a joint scientific policy. The ILT resources have benefitted from the following recent major funding sources: CNRS-INSU, Observatoire de Paris and Université d’Orléans, France; BMBF, MIWF-NRW, MPG, Germany; Science Foundation Ireland (SFI), Department of Business, Enterprise and Innovation (DBEI), Ireland; NWO, The Netherlands; The Science and Technology Facilities Council, UK; Ministry of Science and Higher Education, Poland. TW acknowledges financial support from the grant CEX2021-001131-S funded by MICIU/AEI/ 10.13039/501100011033, from the coordination of the participation in SKA-SPAIN, funded by the Ministry of Science, Innovation and Universities (MICIU). MS and RJD acknowledge funding from the German Science Foundation DFG, within the Collaborative Research Center SFB1491 “Cosmic Interacting Matters – From Source to Signal”. MB acknowledges funding by the Deutsche Forschungsgemeinschaft (DFG, German Research Foundation) under Germany’s Excellence Strategy – EXC 2121 “Quantum Universe” 390833306. This research has made use of the NASA/IPAC Extragalactic Database (NED), which is operated by the Jet Propulsion Laboratory, California Institute of Technology, under contract with the National Aeronautics and Space Administration.  The following software packages have been used in this work: Astropy \citep{astropy_13a,astropy_18a}. 
This research has made use of "Aladin sky atlas" developed at CDS, Strasbourg Observatory, France \citep{bonnarel_00a,boch_14a}; SAOImage DS9 \citep{joye_03a}. This work made use of the SciPy project \href{https://scipy.org}{https://scipy.org}.
\end{acknowledgement}

%
%

\bibliographystyle{aa}
\bibliography{review} 



    \label{fig:stellar}




\end{document}